\documentclass[twocolumn,prl,showpacs]{revtex4}
\usepackage{amstext}
\usepackage{amsmath}            
\usepackage{amssymb}            
\usepackage{graphicx}           
\usepackage{latexsym}

\newcommand{\ket}[1]{|#1\rangle}
\newcommand{\bra}[1]{\langle #1|}
\newcommand{\bk}[1]{\langle #1 \rangle}
\newcommand{\one}{\mbox{$1 \hspace{-1.0mm}  {\bf l}$}}

\begin{document}
\title{Geometric phase in open systems}
\author{ A. Carollo$^{\diamond}$$^\ddag$, I. Fuentes-Guridi$^{\dagger}$,
M. Fran\c{c}a Santos$^{\diamond}$ and  V. Vedral$^{\diamond}$}
\address{$^{\diamond}$Optics Section, The Blackett Laboratory,
Imperial College, London SW7 2BZ, United Kingdom \\
$^{\dagger}$ Perimeter Institute, 35 King Street North Waterloo, Ontario Canada N2J 2W9,\\
$^\ddag$INFM, unit\`{a} di ricerca di Palermo, via Archirafi 36, 90123 Italy}

\begin{abstract}
We calculate the geometric phase associated to the evolution of a system subjected to decoherence through a
quantum-jump approach. The method is general and can be applied to many different physical systems. As examples,
two main source of decoherence are considered: dephasing and spontaneous decay. We show that the geometric phase
is completely insensitive to the former, i.e. it is independent of the number of jumps determined by the dephasing
operator.
\end{abstract}

\pacs{03.65.-w 03.65.Vf 03.65.Yz}   

\maketitle The generalization of the geometric phase~\cite{pac,berry,wil} associated to the evolution of a system
in a mixed state scenario is still an open problem. There have been various proposals tackling the problem from
different perspectives like, for example, via state purification~\cite{uhlm2}, through an interferometric
procedure~\cite{sjoq}, or for dissipative systems~\cite{gamliel}. In most of the cases these definitions do not
agree on account of different constraints imposed, namely different generalizations of the parallel transport
condition.

In this paper, we approach the problem of mixed states geometric phases in open quantum systems through the
quantum jumps method~\cite{qjumpmart,carmic} (for alternative methods see~\cite{peix,rom;pin,eric;sjoq}). We show
that, in general, it is possible to define and calculate a geometrical phase for an open system, which can even be
robust against decoherence, as, for example, it is the case for purely diffusive reservoirs. As an illustration of
our method, we calculate Berry's phases for a spin 1/2 in contact with different reservoirs.

The quantum jumps model proves to be particularly suitable for the case of geometric phases because in each
particular trajectory the quantum state of the system remains pure (if initially pure). The non unitary evolution
of the system and mixed states are recovered when averaging over all possible trajectories. Therefore, we avoid
the problem of finding the proper parallel transport condition, which is very well defined for unitary evolutions.

Let us take a system evolving according to the following general master equation ($\hbar=1$):
\begin{equation}
  \label{eq:mastereq}
  \dot{\rho}=\frac{1}{i}[H,\rho]-\frac{1}{2}\sum_{k=1}^{n}
  \{\Gamma_k^\dagger \Gamma_k\rho+\rho \Gamma_k^\dagger \Gamma_k\ - 2\Gamma_k \rho\Gamma_k^\dagger\},
\end{equation}
For a small time interval $\Delta t$, we can describe the time evolution of the density matrix by
\begin{equation}
  \label{eq:decomp}
  \rho(t+\Delta t)\approx \sum_{k=0}^{n} W_k \rho(t) W_k^\dagger,
\end{equation}
where $W_0=\one-i\tilde{H}\Delta t$ and $W_k=\sqrt{\Delta t} \Gamma_k\quad$ ($k\in\{1\dots n\}$) are called the
``no-jump'' and jump operators respectively. $\tilde{H}$ is a non-Hermitian Hamiltonian, given by:
\begin{equation}
  \label{eq:nonHerHam}
  \tilde{H}=H-\frac{i}{2}\sum_{k=1}^n \Gamma_k^\dagger \Gamma_k.
\end{equation}
Note that the operators $W_k$ fulfill the completeness relation $\sum_{k=0}^nW_k^\dag W_k=\one$.

In this description, the dynamics of the system is approximated by dividing the total evolution time $T$ into a
sequence of discrete intervals $\Delta t = \frac{T}{N}$. According to Eq.~(\ref{eq:decomp}), the state of the
system, after any time step $t_m=m\Delta t$, evolves into $\rho(t_{m+1})=W_k \rho(t_m) W_k^\dagger$ (up to first
order in $\Delta t$), with probability $p_k=Tr{W_k\rho(t_m) W_k^\dagger}$. For example, an initial ($t=0$) pure
state $\psi_0$ would evolve, after the first time interval, into the (not normalized) state $\ket{\psi_0}\to
\ket{\psi_1}=W_k\ket{\psi_0}$ with probability $p_k(t_1)=\bra{\psi_0}W_k^\dag W_k \ket{\psi_0}$.


The time evolution of the system is, then, calculated for a set of possible trajectories containing, each one of
them, different numbers of jumps, occurring at different times, i.e. each trajectory is defined as a chain of
states obtained by the action of a sequence of operators $W_k$ on the initial state. For example, for an initial
pure state $\ket{\psi_0}$, the (non-normalized) state of the system, after the m-th step, along the i-th
trajectory, is given by:
\begin{equation}
\ket{\psi^{(i)}_m}=\prod_{l=1}^{m}W_{i(l)}\ket{\psi_0},
\end{equation}
where $i(l)$ stands for the l-th element of a sequence of indexes with values belonging from ${0\dots n}$. Each
trajectory, then, is represented by a discrete sequence of pure states
$\{\psi_{0},\psi^{(i)}_{0},\dots,\psi^{(i)}_{N}\}$. The dynamics given by the master equation is recovered by
summing incoherently all the states associated to each trajectory, and taking the continuous limit $\Delta t \to
0$.

The fact that a pure state remains pure in each trajectory, in the quantum jumps method, is very useful in our
case, since it is known that, given a chain of pure states $\{\ket{\psi_1}\dots\ket{\psi_N}\}$, the geometric
phase associated to them is given by the Pantcharatnam formula~\cite{muk;sim}:
\begin{equation}\label{eq:defphase}
\gamma_g=-\arg\left\{\bk{\psi_1|\psi_2}\bk{\psi_2|\psi_3}\dots\bk{\psi_{N-1}|\psi_N}\bk{\psi_N|\psi_1}\right\}
\end{equation}
Therefore, we are able to associate a meaningful geometrical phase to each trajectory ``i'' described by the
system, as the continuous limit of Eq.(\ref{eq:defphase}) for the sequence
$\{\psi_{0},\psi^{(i)}_{0},\dots,\psi^{(i)}_{N}\}$.

As an example, let us consider the "no-jump" trajectory for a completely general master equation. The evolution of
a quantum state along this trajectory is obtained by the repeated action of the operator $W_0$. At the time
$t=m\Delta t$, the quantum state will be approximately given by:
\begin{equation}
\label{eq:nojumpcontk}
 \ket{\psi^{0}_m}=(W_0)^{m}\ket{\psi_0}=\left(\one-i\frac{T}{N}\tilde{H}\right)^{\frac{N}{T}t}\ket{\psi_0}.
\end{equation}
which in the continuous limit $N\to \infty$ yields to a dynamics governed by the complex effective Hamiltonian
$\tilde{H}$:
\begin{equation}
\label{eq:nojumpcont}
i\frac{d}{dt} \ket{\psi^{0}(t)}=\tilde{H}\ket{\psi^{0}(t)}\quad \ket{\psi^0(0)}=\ket{\psi_0}
\end{equation}
Thus, the evolution corresponding to this trajectory is given by a smooth chain of (non normalized) states
$\ket{\psi(t)}$, in which case $\gamma_N$ converges to:
\begin{equation}\label{eq:gpnojump}
\gamma=-Im \int_0^T \frac{\bra{\psi(t)}\frac{d}{dt}\ket{\psi(t)}}{\bk{\psi(t)|\psi(t)}} dt -
\arg\{\bk{\psi(T)|\psi(0)}\}.
\end{equation}
Substituting Eq.~(\ref{eq:nojumpcont}) into Eq.(\ref{eq:gpnojump}), we obtain the geometric phase for a no-jump
trajectory, which is given by:
\begin{equation}\label{eq:gpnojump2}
  \gamma^0=\int_0^T \frac{\bra{\psi^0(t)}H\ket{\psi^0(t)}}{\bk{\psi^0(t)|\psi^0(t)}} dt -
\arg\{\bk{\psi^0(T)|\psi^0(0)}\}
\end{equation}
This is the geometric phase associated to a non-unitary evolution of a system~\cite{pati,muk;sim}, when there are
no jumps. The first term is clearly the opposite of the dynamical phase associated to the non-unitary evolution,
as it is given by the average of the Hamiltonian (up to a minus sign) along the path traversed by the system. The
second term is the total phase difference between the final and the initial state, according to Pancharatnam's
definition of distant parallelism~\cite{pac}. Thus the geometric phase is obtained as the difference between total
and dynamical phase associated to a given evolution of pure states~\cite{muk;sim}.

Note that, in the special case in which $\sum_{i=1}^n \Gamma_i^\dag \Gamma_i\propto \one$ (which is a unital
evolution), the geometric phase associated with the no-jump trajectory is the same as the one acquired by an
isolated system evolving under the same Hamiltonian $H$. This becomes clear when one notes that, in this case,
$W_0=(1-\alpha)\one+iH\Delta t$ and the evolution of state $\ket{\Psi(t)}$ is the same as its isolated counterpart
up to a global normalization factor $e^{-\alpha t}$. In other words, for this particular source of decoherence, if
the reservoir is permanently measured and no jump is detected, there is no gain of information on the system,
which simply projects it back into its unitary evolution.

Note, also, that following the idea of~\cite{sam;ban}, it is possible to represent the geometric
phase~(\ref{eq:gpnojump2}) as the integral of the Berry connection form:
\begin{equation}
  \label{eq:berryconn}
  d\omega=Im\frac{\bra{\psi(t)}d\ket{\psi(t)}}{\bk{\psi(t)|\psi(t)}}
\end{equation}
along a closed path. This path is formed by the trajectory $\psi(t)$ followed by the states along the Hilbert
space during the dynamical evolution and the shortest geodesic connecting final and initial states $\psi(T)$ and
$\psi(0)$. Thus the second term of equation~(\ref{eq:gpnojump2}) can be regarded as the path integral of the Berry
connection along this geodesic.


Suppose, now, that there is only one jump in the trajectory at an arbitrary time $t_1$, which occurs in a time
much shorter than any other characteristic time of the system. Then, we can separate the evolution in two parts
(before and after the jump) and, the continuous limit of equation~(\ref{eq:defphase}) leads to the following
expression:
\begin{eqnarray}
\label{eq:gp1jump}
    \gamma^1_j&=&\int_0^{t_1} \frac{\bra{\psi^{'}(t)}\frac{d}{dt}\ket{\psi^{'}(t)}}{\bk{\psi^{'}(t)|\psi^{'}(t)}} dt -
\arg\{\bk{\psi^{'}(t_1)|\Gamma_j|\psi^{'}(0)}\}+ \nonumber \\
&&\int_{t_1}^{T} \frac{\bra{\psi^{''}(t)}\frac{d}{dt}\ket{\psi^{''}(t)}}{\bk{\psi^{''}(t)|\psi^{''}(t)}} dt -
\arg\{\bk{\psi^{''}(T)|\psi^{'}(0)}\}
\end{eqnarray}
where $W_j$ is the operator associated to the occurred jump, and $\psi^{'}(t)$ and $\psi^{''}(t)$ are the states
evolving under the effective Hamiltonian $\tilde{H}$, before and after the jump respectively. They are given by
the equation~(\ref{eq:nojumpcont}) with initial conditions $\psi^{'}(0)=\psi_0$ and
$\psi^{''}(t_1)=W_j\psi^{'}(t_1)$, respectively.

The first and third term represents the dynamical phase given by the effective evolution~(\ref{eq:nojumpcont}),
before and after the jump occurs. The last term is the phase difference between initial and final state of the
total evolution. The second term is a phase associated to the occurrence of a jump at time $t_1$. Analogously to
the total phase associated to final and initial state, this term represents the phase difference between the
states after and before the jump, and geometrically, it can be regarded as the path integral of the Berry
connection along the shortest geodesic joining them.

This result can be easily generalized to any trajectory, allowing for a more complicate sequence of jumps and
no-jump evolutions. The geometric phase is then represented as the sum of terms of the form
$\arg{\bk{\psi(t_i)|\Gamma_{j}|\psi(t_i)}}$ regarded as the phase associated to the jump $\Gamma_{j}$ occurring at
the instant $t_i$, and terms of the form~(\ref{eq:gpnojump}) for the no jump evolutions. And clearly all these
phases can be regarded as the integrals of Berry connection along a complex path composed of geodesics joining
initial and final state of the jumps, and the paths traversed by the state during the evolution under $\tilde{H}$.

Let us apply this general quantum jumps procedure to a well known physical system. First, let us consider the
simplest example of decoherence: a two levels system evolving under the free Hamiltonian
$H=\frac{\omega}{2}\sigma_z$ and subjected to dephasing, which can be described by the Master
equation~(\ref{eq:mastereq}) with $\Gamma=\lambda\sigma_z$, where $\lambda$ is the the coefficient giving the
probability per unit time of a ``phase-jump''.

Since this is a decoherence model for which $\Gamma^\dagger\Gamma=\sigma_z^2 \propto \one$, which is a simple
instance of a unital evolution, according to the previous considerations, the geometric phase associated to the
no-jump case is given by the standard geometric phase associated to the unitary evolution of a spin 1/2 linearly
coupled to a constant magnetic field. For instance, after a time $t=2\pi/\omega$,
$\gamma_0=\pi(1-\bk{\psi_0|\sigma_z|\psi_0})=\pi(1-\cos \theta)$, where $\psi_0$ is the initial state and $\theta$
is its azimuthal angle in the Bloch sphere representation.

Although the no-jump case may seem trivial, this system has a much more remarkable property: the geometric phase
is actually robust against dephasing, in this simple, but very useful example. In fact, we show below that the
final geometric phase is unaffected by any number of jumps for any particular trajectory. To show that, Let us
consider first the case of a single jump, in which the phase is given by:
\begin{eqnarray*}
  \label{eq:gpsz1jmp}
\gamma_1&=&-\int_0^{t_1} \frac{\omega}{2} \bra{\psi_0}\sigma_z\ket{\psi_0} dt -
\arg\{\bk{\psi_0|\sigma_z|\psi_0}\}\\
&-&\int_{t_1}^{2\pi/\omega} \frac{\omega}{2} \bra{\psi_0}\sigma_z\ket{\psi _0} dt \\
&-& \arg\{\bk{\psi_0|e^{i\frac{\sigma_z}{2}(2\pi-\omega t_1)}\sigma_z e^{i\frac{\sigma_z}{2}\omega t_1}|\psi_0}\}\\
&=&\pi(1-\bk{\psi_0|\sigma_z|\psi_0})=\pi(1-\cos \theta),
\end{eqnarray*}
where the fact that $H$ and $\Gamma$ commute has been used. This result is easily generalized to any number $k$ of
jumps:
\begin{eqnarray*}
\gamma_k&=&-\int_0^{2\pi/\omega} \frac{\omega}{2} \bra{\psi_0}\sigma_z\ket{\psi_0} dt -
\arg\{\bk{\psi_0|\sigma_z|\psi_0}^k\}-\\
&-& \arg\{\bk{\psi_0|e^{i \sigma_z \pi}(\sigma_z)^k|\psi_0}\}=\pi(1-\cos \theta),   \label{eq:szkjmp}
\end{eqnarray*}
Thus, no matter how many jumps occur in the chosen trajectory, we can associate the same geometric evolution to
the system. There is a simple geometrical explanation for this effect. Dephasing is a special source for
decoherence because it does not change the projection of the spin vector on the direction of the magnetic field,
i.e. it does not change the relative angle $\theta$ between the directions of the magnetic field and the spin.
After each jump, the spin is still precessing around the magnetic field alongside the same curve. As a result, the
total area covered by its trajectory remains the same, and so does the geometric phase acquired by the spin state,
which is proportional to this area. Therefore, in the end, the geometric phase acquired by the spin state will be
the same, no matter how diffused its total phase may be. That does not mean that dephasing will not affect the
measurement of this phase. Indeed, it will lower the visibility of any interference measurement made on the spin,
because the visibility of the state is lowered when its mixedness is increased (we will address this in more
details in a separate publication). However, as the calculations above show, the reduced visibility will be caused
by a randomization of the dynamical phase, and not the geometrical one, which proves to be much more robust in
this case.

A more realistic example includes spontaneous decay as a source of decoherence for the spin 1/2 system. In this
case, it is only worth analyzing the no-jump case, since any jump causes immediate and complete loss of phase
information of the quantum state. Spontaneous decay $\Gamma=\alpha \sigma_-$ is a decoherence source that cannot
be associated to a unital map ($\sigma_+\sigma_-\neq \one$) and, therefore, the phase will be affected even if no
jump is detected. However, as we show in figure 2, the no-jump trajectory is a smooth spiral converging to the
lower state, which still allows us to calculate the phase using Eq.~(\ref{eq:gpnojump}). We obtained
$\gamma=\pi+\frac{\omega}{2\alpha} \ln \left(\bk{\psi_0|e^{-4\pi \frac{\alpha}{\omega} \sigma_z}|\psi_0} \right)$,
which in the limit $\omega\gg \alpha$ leads to
\[
\gamma\approx\pi(1-\cos\theta)+(4\pi)^2\frac{\alpha}{\omega} \sin^2{\theta} +o\left(\frac{\alpha}{\omega}\right)^2
\]
Again, this result has a very simple geometrical explanation: as we observe the reservoir and detect no jump, the
probability that the system is in the lower state smoothly increases, changing $\theta$ and, therefore, the
element of area covered by the spin trajectory in each infinitesimal time interval, as shown in figure 2.

Another simple case that can be analyzed is the spin flip alongside an arbitrary direction
$\Gamma=\sigma_{\hat{n}}$. In this case, the no jump situation is again trivial and similar to the dephasing
reservoir, since $\sigma_{\hat{n}}^2=\one$. When one or more jumps occur, we can use Eq.(\ref{eq:gp1jump}) (or its
generalization to many jumps) to easily calculate the final phase, which will be a sum of the partial areas
covered in each trajectory with plus or minus sign depending on the respective coupling energy of the spin with
the magnetic field. Our treatment is, of course, applicable even when the master equation contains many different
sources of errors acting simultaneously on the system, since we can use the generalized form of
equation~(\ref{eq:gp1jump}) to calculate the phase.

In conclusion, in this paper, we present a method to calculate geometric phases in open systems. Our method is
general and can be applied as long as the system dynamics is described by a master equation in the form of
Eq.(\ref{eq:mastereq}), which is the most general completely positive trace preserving continuous
evolution~\cite{lindblad}. By using the quantum jumps approach we avoid the problem of defining Berry's phases for
mixed states: in each trajectory, the quantum state of the system remais pure and the phase can be calculated
through usual procedures. In particular, we show that it is always possible to calculate this phase, either for
the no-jump trajectories or for the ones in which one or more jumps occur. We also show that, for special unital
decoherence sources, the phase remains unaffected for the no jumps trajectories. As a direct application of our
method, we calculate the geometric phases of spin 1/2 systems coupled to different reservoirs. We show that those
phases are totaly robust against phase diffusion, in which case the lower visibility observed due to the
non-unitary evolution may be attributed solely to a randomization of the dynamical phase. This property may be
interesting for possible applications, specially in quantum computing, since dephasing may be difficult to monitor
and correct, in general. Therefore, it is interesting noticing that geometric phases are robust against this
decoherence source. We also present a nice geometrical explanation to this effect, as well as to the effect on the
geometric phase when spontaneous emission is present, but no jump is detected. We also briefly comment on other
typical decoherence effects on the system, like arbitrary spin flips. The method presented here is completely
general and can be applied to many other physical systems.

This research was supported by EPSRC, Hewlett-Packard, Elsag spa and the EU and Quiprocone Grant No. 040. MFS
acknowledges the support of CNPq.

\begin{figure}
\includegraphics[width=5cm]{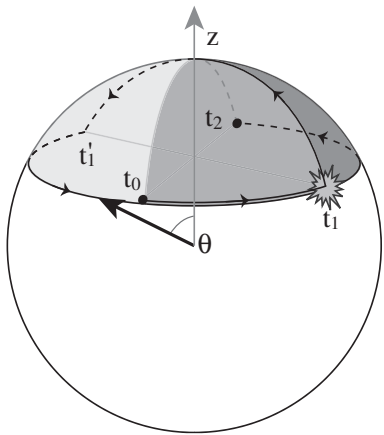}
\caption{Evolution of the state along a "one-jump" trajectory on the Bloch sphere, in the case of phase diffusion
decoherence ($\Gamma\propto \sigma_z$). From time $t_0$ to time $t_1$ the state evolves under the no-jump
hamiltonian $\tilde{H}$ alongside the parallel of the sphere. At time $t_1$ a jump occurs, flipping
(instantaneously) the Bloch vector about the $z$ to the point $t_1^\prime$, and the no-jump evolution starts
again. At time $t_2=t_0+2 \pi/\omega$ the geometric phase $\gamma=\pi(1-\cos\theta)$ is recovered. The geometric
phase is half the area enclosed in the path spanned by the Bloch vector. This is given by two contribution, the
surface $t_1-t_0-t_1^\prime$ and the surface $t_0-t_1-t_2$. } \label{fig:dephase}
\end{figure}

\begin{figure}
\includegraphics[width=5cm]{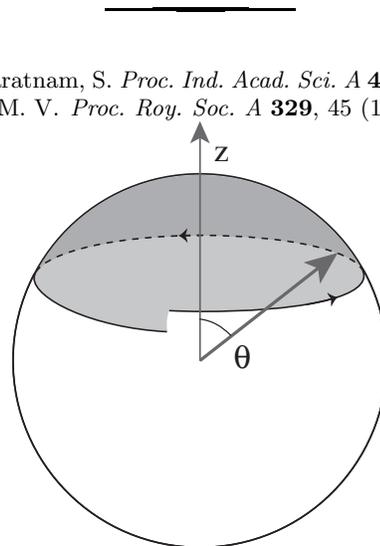}
\caption{Evolution of the state along the no-jump trajectory on the Bloch sphere, for the spontaneous decay case
($\Gamma_1=\lambda \sigma_z$, $\Gamma_2=\alpha \sigma_-$). The evolution is a smooth spiral converging to the
lower state: while the state rotates about the $z$ axis, it is smoothly brought towards the lower state with a
velocity given by the spontaneous decay rate $\alpha$. The geometric phase is given by the area enclosed in the
path shown, where the last segment is the geodesic connecting initial and final point of the evolution.
}\label{fig:spdecay}
\end{figure}

\bibliographystyle{nature}
\bibliography{decoh}

\end{document}